# Quantifying Data Rate and Bandwidth Requirements for Immersive 5G Experience

Yinan Qi[1], Mythri Hunukumbure[1], Maziar Nekovee[1], Javier Lorca[2], and Victoria Sgardoni[3]

[1]Samsung Electronics R&D Institute UK, Staines, Middlesex TW18 4QE, UK
[2]Telefónica I+D – GCTO, C/ Zurbarán 12, 28010 Madrid, Spain
[3]Department of Electrical and Electronic Engineering, University of Bristol, Bristol BS8 1UB, UK
[1]yinan.qi, mythri.h, and m.nekovee@samsung.com, [2]franciscojavier.lorcahernando@telefonica.com,
[3]Victoria.Sgardoni@bristol.ac.uk

*Abstract*—The proliferation of smartphones/mobile devices that support a wide range of broadband applications and services has driven the volume of mobile data traffic to an unprecedented high level, requiring a next generation mobile communication system, i.e., the fifth generation (5G). Millimeter wave bands, due to the large available spectrum bandwidth, are considered as one of the most promising approaches to significantly boost the capacity. In this paper, we define a typical use case envisaged in the early stage of the 5G system rollout, where users can experience 100M+ data rate in the target area and at the same time enjoy services demanding extremely high data rates, such as virtual reality and ultra-high definition video. We then break down the use case into four different traffic types: web browsing, content sharing, virtual reality experience and ultra-high definition video, and derive and analyze the distributions of the required instantaneous data rates for each individual traffic type. Finally, we consider the case where multiple users with mixture of traffic types are simultaneously active and analyze the overall data rate and bandwidth requirements to support such a scenario.

*Keywords—5G; extreme data rate, immersive experience, millimeter wave, small cells, usage distribution functions*

## I. INTRODUCTION

Mobile communications operating in higher millimeter-wave (mm-wave) frequencies are being seriously considered by industry as a very promising approach to significantly boost the capacity of the next generation systems (5G). Indeed, such a system can potentially utilize the much larger spectrum bandwidth available in mm-wave frequency bands and support Gbps data rates in a commercially viable manner. Contiguous bandwidths larger than 100 MHz and up to several GHz are in principle available for future mobile usage (at least on a co-primary basis) in this frequency range. For simplicity, in this paper we refer to "mm-wave spectrum" as frequencies in the range 6-100 GHz, which so far have been largely unexplored for cellular use.

The European Commission funded mmMAGIC project [1]-[2] investigates the potential suitability of the mm-wave spectrum for mobile operations and also defines radio architectures, design concepts and solutions in this frequency range. As part of the project, 5G use cases (with potential *key performance indicator* (KPI) requirements) have been developed and the bandwidth needs to achieve those KPIs have been analyzed.

In this paper, we focus on one such use case developed in the mmMAGIC project [3]. This use case, entitled 'Immersive early 5G experience in targeted coverage', envisions to provide a distinct 5G experience to the early 5G adopters in targeted areas like hotspots. We believe satisfying or even exceeding the expectations of these early adopters will be crucial to the success of 5G, at least in the mobile communications sphere. In section II of this paper, a full description of the use case is provided. Section III is devoted to analyzing the data rate requirements of each of the four data types demanded by the early 5G users defined in this use case. Some numerical results are presented in section IV, with an overall bandwidth requirement analysis. The paper is concluded in section V.

## II. USE CASE DESCRIPTION

Unlike in the dawn of previous mobile generations, it is very likely that 5G will usher in new services from its inception. The research into 5G today should develop KPIs for these kinds of services/use cases and the technology to support these KPIs. As a pre-standardisation research project, mmMAGIC places special emphasis on identifying such use cases and developing the mm-wave technical solutions to realize them. The 'Immersive early 5G experience in targeted coverage' use case was developed with this intention and with the view that meeting or even exceeding early 5G adaptors' usage requirements will be critical to the accelerated expansion of 5G services.

The early 5G adopters we are targeting in this use case would particularly want to embrace the immersive multi-media experiences including 4k/8k UHD video, virtual reality experiences and real time mobile gaming. These 5G experiences should come with a distinct improvement in Quality of Experience (QoE) compared to that in the (then) legacy 4G services. For the immersive multi-media experiences [4], the 'user experienced' data rates, latencies and other key KPIs should indicate a step change from the (then) 4G evolutions.

In determining the key KPIs for this use case, we have to consider the likely requirements of the evolving immersive applications and the likely capabilities that the evolving 4G systems would provide. Currently the 4k UHD video streaming service promoted in UK needs at least 40Mbps consistent data-rates [5], which are today only possible with wired/satellite connections. With 8k UHD and further evolutions these requirements are likely to increase many fold. The current LTE-Advanced specifications can provide around 10Mbps user experienced data rates and 1Gbps peak data rates for static users. The latencies supported by these specs are around 50ms. The proposed key KPIs for this 5G use case are: 100Mbps as a baseline data rate, while the peak data rates (on demand) can be up to 20Gbps and the latencies to be below 10ms. Such a large variance in data rates comes from the intention to support the potential requirements of the evolving immersive 5G applications. The 20Gbps peak rate can seem excessive at first glance. However, rival technologies like WiGig are already proposing peak rates of around 7Gbps [6]. New services will develop in this context and 5G mobile data needs to stay competitive with such extreme data rates.

Reliability will also be a crucial KPI of the high QoE expected by the early 5G adopters, especially for UHD video streaming and evolving immersive 5G services. In addition, these initial 5G hotspots, deployed on mm-wave spectrum, will have an underlay coverage provided by 3G, 4G or even sub 6GHz 5G systems. The interworking, handover co-ordination between these 5G hotspots and the underlay network will be an important feature for this use case.

As noted above, the user experienced data rate should improve 10 times and the peak data rates should improve 20 times from the current IMT-Advanced levels. The latencies should be reduced by at least 5 times to achieve (near) real time experience in immersive 5G services. One of the deployment challenges compared to LTE would be the densification of the small cells. We believe that around 25 small cells per hotspot area (typically 0.1km$^2$) would be required to provide the necessary capacity and the number of connections. Interference control and back-haul provision technologies will also have to significantly improve from current state-of-the-art.

III. USE CASE ANALYSIS

The data rate requirements for a 5G user are expected to fluctuate dramatically depending on different service types. Accurate prediction of a 5G user's behavior may not be feasible, but a statistical traffic model will help operators to understand and approximate the user's behavior in general as well as in extreme cases, where very high instantaneous data rates can be demanded. To meet very high throughput requirements, operators need to focus on the instantaneous data rates rather than on the average data rates. We assume that the traffic model of a 5G user can be generated based on a mixture of four typical traffic types: 1) web browsing, where users do web surfing, online shopping, etc.; 2) content sharing, where users share pictures, documents or short videos from other users; 3) virtual reality experience, where very high data rate information exchange is expected in stationary/low mobility environment; and 4) UHD video, where UHD (4K/8K) is expected to be available on mobile devices. In this section, we will introduce traffic patterns and derive the distribution functions of the required instantaneous data rate for each individual traffic type and then combine all the traffic models to give a holistic model and analysis for the proposed 5G use case.

A. Web Browsing

The web browsing session can be divided into ON/OFF periods, where the web pages are downloaded as data packets during ON periods and OFF periods represent reading time identifying the required time by the users to digest the web pages. The packet size $X$ is a random variable and assumed to follow the Truncated Lognormal Distribution [7]. The probability density function (pdf) of $X$ is given as

$$f(x) = \frac{1}{\sqrt{2\pi}\sigma} \exp\left(-\frac{(\ln x - \mu)^2}{2\sigma^2}\right), -\infty < a_{low} \leq x \leq a_{up} < \infty \quad (1)$$

where $a_{low}$ and $a_{up}$ identify the lower and upper limits of the most prominently used packet size for web browsing. Assuming another random variable $Y=\ln X$, the pdf of $Y$ can be expressed as

$$f(y) = \frac{\frac{1}{\sigma}\varphi\left(\frac{y-\mu}{\sigma}\right)}{\Phi\left(\frac{\ln a_{up} - \mu}{\sigma}\right) - \Phi\left(\frac{\ln a_{low} - \mu}{\sigma}\right)}, \ln a_{low} \leq y \leq \ln a_{up} \quad (2)$$

where $\Phi(y)$ is the Gaussian error function and $\varphi(y)$ is the function of the standard normal distribution

The reading time, i.e., inter-arrival time (IAT) $T$ is also a random variable and is assumed to follow exponential distribution [7] as

$$f(t) = \frac{1}{T_{wb}} \exp\left(-\frac{1}{T_{wb}} t\right), 0 \leq t < \infty \quad (3)$$

where $T_{wb}$ is the mean IAT. Since the IAT is normally large when a person is browsing a website, and the packet size is expected to be small, we can assume that a web page packet can be scheduled and transmitted before the arrival of the next packet for simplicity. Thus the instantaneous data rate in such a case can be given as

$$R_{wb} = \frac{X}{T} \quad (4)$$

The distribution of $R_{wb}$ can be defined as a ratio distribution, which can be derived from the joint distribution of the two independent random variables, e.g., $Y$ and $T$, by integration of the following form [8]

$$f(r_{wb}) = \int_0^{+\infty} t p_{Y,T}(r_{wb}t, t) dt \quad (5)$$

where $p_{Y,T}$ is the joint distribution of $Y$ and $T$, given as

$$p_{Y,T}(y,t) = \frac{1}{T_{wb}} \frac{\frac{1}{\sigma}\varphi\left(\frac{y-\mu}{\sigma}\right)\exp\left(-\frac{1}{T_{wb}}t\right)}{\Phi\left(\frac{\ln a_{up} - \mu}{\sigma}\right) - \Phi\left(\frac{\ln a_{low} - \mu}{\sigma}\right)} \quad (6)$$

$$\ln a_{low} \leq y = \ln x \leq \ln a_{up}, 0 \leq t < \infty$$

In Appendix A, (5) is derived as

$$f(r_{wb}) = \overline{K}\left(\overline{\mu} + \frac{\overline{\sigma}\left(\varphi\left(\frac{(\ln a_{up})/r_{wb} - \overline{\mu}}{\overline{\sigma}}\right) - \varphi\left(\frac{(\ln a_{low})/r_{wb} - \overline{\mu}}{\overline{\sigma}}\right)\right)}{\Phi\left(\frac{(\ln a_{up})/r_{wb} - \overline{\mu}}{\overline{\sigma}}\right) - \Phi\left(\frac{(\ln a_{low})/r_{wb} - \overline{\mu}}{\overline{\sigma}}\right)}\right) \quad (7)$$

where $\overline{K}$, $\overline{\mu}$ and $\overline{\sigma}$ are functions of $r_{wb}$ and defined in Appendix A.

### B. Content Sharing

For content sharing, a traffic model similar to FTP traffic model defined in 3GPP is employed [7] and [9]. The packet size is assumed to be fixed as $S_{cs}$. The IAT $T$ is assumed to follow exponential distribution with mean IAT $T_{cs}$. However, the IAT for content sharing is normally much smaller than that of web browsing. In this regard, we assume that $N$ packets will be scheduled and transmitted together. The total duration of $N$ packets is therefore the summation of $N$ identical independent distributed (i.i.d.) random variables as

$$T = \sum_{i=1}^{N} T_i \quad (8)$$

where each $T_i$ follows the same exponential distribution. According to [10], the random variable $T$ follows the Erlang distribution as

$$f(t) = \frac{\lambda_{cs}^N t^{N-1}}{(N-1)!} \exp(-\lambda_{cs} t), 0 \leq t < \infty \quad (9)$$

where $\lambda_{cs} = 1/T_{cs}$. The instantaneous data rate required by content sharing traffic therefore can be expressed as

$$R_{cs} = \frac{NS_{cs}}{T} \quad (10)$$

The probability of the random variable $R_{cs}$ being smaller than $r_{cs}$ can be easily calculated as

$$\Pr(R_{cs} < r_{cs}) = \sum_{n=0}^{N-1} \frac{1}{n!} \exp\left(-\lambda_{cs} \frac{NS_{cs}}{r_{cs}}\right)\left(\lambda_{cs} \frac{NS_{cs}}{r_{cs}}\right)^n \quad (11)$$

The distribution of $R_{cs}$ can be obtained as

$$f(r_{cs}) = \frac{\partial \Pr(R_{cs} < r_{cs})}{\partial r_{cs}} = \frac{(NS_{cs}\lambda_{cs})^N}{(N-1)! r_{cs}^{N+1}} \exp\left(-\lambda_{cs} \frac{NS_{cs}}{r_{cs}}\right), 0 \leq r_{cs} < +\infty. \quad (12)$$

where $\lambda_{cs} = 1/T_{cs}$.

### C. VR Experiences

The VR experience traffic can be modelled in a similar way as the content sharing traffic but with larger packet size $S_{vr}$ and a smaller mean IAT $T_{vr}$. The distribution of the instantaneous rate for VR is therefore given as

$$f(r_{vr}) = \frac{\partial \Pr(R_{vr} < r_{vr})}{\partial r} = \frac{NS_{vr}\lambda_{vr}^N r_{vr}^{N-3}}{(N-1)!} \exp(-\lambda_{vr} r_{vr}), 0 \leq r_{vr} < +\infty \quad (13)$$

where $\lambda_{vr} = 1/T_{vr}$.

### D. UHD Video

Various aspects of UHD television such as display resolution, frame rate, etc. are defined in [11]. It allows for UHD TV frame rates of 120p, 60p, 59.94p, 50p, 30p, 29.97p, 25p, 24p, and 23.976p. The following table illustrates different image types [12].

Table-1 Image Types

| Bits Per Pixel | Number of Colors Available | Common Name(s) |
|---|---|---|
| 1 | 2 | Monochrome |
| 2 | 4 | CGA |
| 4 | 16 | EGA |
| 8 | 256 | VGA |
| 16 | 65536 | XGA, High Color |
| 24 | 16777216 | SVGA, True Color |
| 32 | 16777216 + Transparency | |

The "bits per pixel" (bpp) refers to the sum of the bits in all three color channels and represents the total colors available at each pixel. Based on these values, we can do the following calculation for the required average data rate with different frame rate as

$$R_{ave} = bpp \times resolution \times R_{fm} \quad (14)$$

where resolution is 3840×2160 and 7680×4320 for 4K and 8K UHD video, respectively, and $R_{fm}$ is the frame rate.

Without employing any advanced video coding scheme the uncoded data rate can vary between 3.182Gbps up to 127.4Gbps, which is a very wide range. As aforementioned, it is expected that 5G systems will provide up to 20Gbps data rate. In this regard, most of the frame rates of 4K UHD can be supported. Actually, most of the frame rates result in data rates between 3-10Gbps and more than 10Gbps data rate is only needed for four settings. For 8K UHD, without video coding employed, only 16 bpp mode is supported and the data rates vary between 10-20Gbps.

If advanced video coding scheme is considered, e.g., H.264, which is currently one of the most commonly used formats for the recording, compression, and distribution of video content and is supposed to achieve 50% data rate reduction [13], some frame rates of the 8K UHD with 32 bpp can be supported by 5G systems, requiring a data rate ranging from 10 to 20Gbps. It is expected that new video coding schemes, e.g., HEVC, can achieve further 40% data rate reduction [5], thus more frame rate options for 8K UHD with 32 bpp can be supported except the 120p one. The overall data rate range is from 0.9546Gbps to 19.11Gbps. However, it should be noted that both H.264 and HEVC are lossy coding.

We assume 4K UHD video will be supported in the early stage of 5G and take 4K UHD, 32bpp, 60p as an example. For UHD, the size of each data frame is the same as $S_F$ and the duration of each frame is also fixed, so that a constant video streaming rate can be achieved. However, within each frame, one large data packet will be divided into a fixed number of data packets and the size of each packet follows the truncated Pareto distribution [7] as

$$f(x) = \frac{\alpha_x a_{x,low}^{\alpha_x} x^{-\alpha_x - 1}}{1 - \left(\frac{a_{x,low}}{a_{x,up}}\right)^{\alpha_x}} \quad (15)$$

where $-\infty < a_{x,low} \leq x \leq a_{x,up} \leq +\infty$ and $\alpha_x > 0$.

The IAT also follows the truncated Pareto distribution [7] as

$$f(t) = \frac{\alpha_t a_{t,low}^{\alpha_t} x^{-\alpha_t - 1}}{1 - \left(\frac{a_{t,low}}{a_{t,up}}\right)^{\alpha_t}} \quad (16)$$

where $-\infty < a_{t,low} \leq t \leq a_{t,up} \leq +\infty$ and $\alpha_t > 0$. Similar to the web browsing case, the instantaneous data rate required by UHD video traffic is a ratio distribution, expressed as

$$f(r_{uhd}) = \int_0^{+\infty} t p_{X,T}(r_{uhd} t, t) dt \quad (17)$$

where $p_{X,T}$ is given as

$$p_{X,T}(x,t) = \frac{\alpha_x a_{x,low}^{\alpha_x} x^{-\alpha_x - 1}}{1 - \left(\frac{a_{x,low}}{a_{x,up}}\right)^{\alpha_x}} \frac{\alpha_t a_{t,low}^{\alpha_t} t^{-\alpha_t - 1}}{1 - \left(\frac{a_{t,low}}{a_{t,up}}\right)^{\alpha_t}} \quad (18)$$

with $-\infty < a_{x,low} \leq x \leq a_{x,up} \leq +\infty$, $-\infty < a_{t,low} \leq t \leq a_{t,up} \leq +\infty$, $\alpha_x > 0$ and $\alpha_t > 0$. In Appendix B, (21) is derived as

$$f(r_{uhd}) = \frac{\overline{K} a_{\min}^{\overline{\alpha}}}{1 - \left(\frac{a_{\max}}{a_{\min}}\right)^{\overline{\alpha}}} \left(\frac{\overline{\alpha}}{\overline{\alpha} - 1}\right) \left(\frac{1}{a_{\min}^{\overline{\alpha} - 1}} - \frac{1}{a_{\max}^{\overline{\alpha} - 1}}\right) \quad (19)$$

where $\overline{K}$, $\overline{\alpha}$, $a_{\min}$ and $a_{\max}$ are defined in Appendix B and $\overline{K}$, $a_{\min}$ and $a_{\max}$ are functions of $r_{uhd}$.

### E. Overall Traffic Model

As aforementioned, a single user could have very diverse service requirements from very low data rate for web browsing to extremely high data rate for UHD video. In this regard, we define the engaging rate to indicate the percentage of time for one user engaging in a specific service. The overall distribution of the traffic model can be expressed as

$$f(r) = p_{wb} f(r_{wb}) + p_{cs} f(r_{cs}) + p_{vr} f(r_{vr}) + p_{uhd} f(r_{uhd}) \quad (20)$$

where $p$ is the engaging rate.

There are multiple active users in each mm-wave small cell [14], the overall traffic demand is therefore the summation of the rate from all users. If we assume there are $N_{ue}$ active users per each mm-wave small cell, the overall data rate $R$ is also a random variable and can be expressed as

$$R = \sum_{i=1}^{N_{ue}} R_i \quad (21)$$

where $R_i$ is the data rate of each user, modelled as an independent and identically distributed (i.i.d.) random variable following the distribution in (24). According to [15], the summation of multiple i.i.d. random variables is given as

$$f_R(r) = (f_{R1} * f_{R2} \cdots f_{RN_{ue}})(r) \quad (22)$$

where $f_{Ri}*f_{Rj}$ represent the convolution of $f(R_i)$ and $f(R_j)$. As we can see from (20), the distribution function of $R_i$ is already very complicated and thus the $N_{ue}$-fold convolution in (22) is very difficult to calculate, even with numerical integration. However, we can also resort to Monte-Carlo method to generate and evaluate the overall data rate $R$.

Spectral efficiency 7.3bits/Hz can already be achieved with some emerging technology [16]. We assume that the current areaspectral efficiency can be enhanced four times due to beamforming and cell densification thereby achieving 7.3×4 bits/Hz area spectral efficiency, under favorable conditions. Therefore the required bandwidth can be easily obtained from the user data rate as;

$$B = R/(7.3 \times 4) \quad (23)$$

## IV. NUMERICAL RESULTS

In this section, we present numerical results for data rate and bandwidth analysis. The parameters are defined in Table-2. Monte Carlo simulation is conducted. For a single traffic, $10^6$ random values are generated based on the distribution of the traffic and PDF of these random values are presented based on histogram. For multiple traffic scenarios, $N_{ue}$ active users are assumed for each simulation run and each individual user's traffic is randomly chosen based on the engaging rate $p$ in (20). Totally, $10^6$ simulation runs are carried on.

Table-2 Parameters

| Parameters | Value |
|---|---|
| Web browsing traffic model | $a_{low}$=100 bytes, $a_{up}$=2 Mbytes, $\sigma$=1.37, $\mu$=8.35, $T_{wb}$=30s |
| Content sharing traffic model | $S_{cs}$=2 Mbytes, $\lambda_{cs}$=8.33 |
| VR experiences | $S_{vr}$=20 Mbytes, $\lambda_{cs}$=50 |
| UHD video (4K, 60p) | $a_{x,up}$=20.75Mbps, $a_{x,low}$=3.32Mbps, $\alpha_x$=1.67, $a_{t,up}$=5.2ms, $a_{t,low}$=0.832ms, $\alpha_t$=1.67, |
| Engaging rate | $p_{wb}$=0.51, $p_{cs}$=0.45, $p_{vr}$=0.02, $p_{uhd}$=0.02 |
| $N_{ue}$ | 40 |

## A. Individual Distribution Functions

The analytical distribution functions are plotted for web browsing, content sharing and UHD video and compared with results obtained via Monte Carlo simulation in Fig. 1 to Fig. 3.

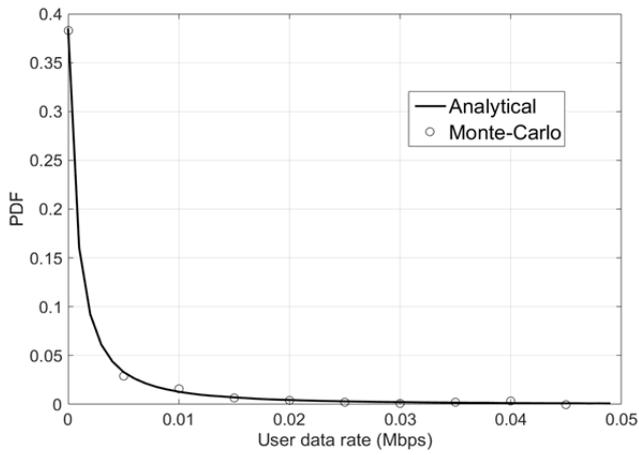

Fig. 1. Pdf of Web browsing required data rate

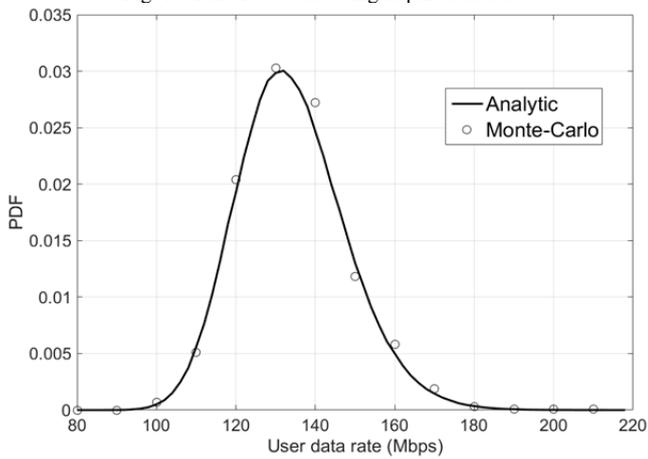

Fig. 2. Pdf of Content sharing required data rate

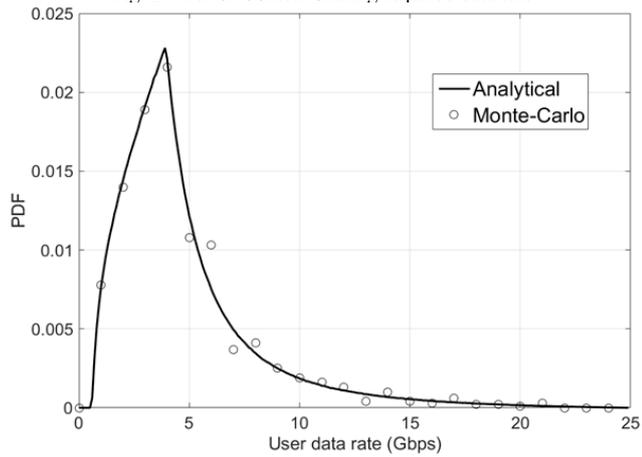

Fig. 3. Pdf of UHD video required data rate

Good matching between the two sets of curves is observed, which proves the accuracy of the derived pdfs in section III.

As can be seen, the required user data rate of web browsing is very small (less than 0.05Mbps). However, it is significantly increased in content sharing, where the mean of the required user data rate is around 135 Mbps and the maximum rate can be up to over 200 Mbps. The data rate is further increased in VR experience and UHD video traffic, where the means are 8 Gbps and 4 Gbps, respectively. Even though the mean data rate of UHD video traffic is lower than that of VR experience, its maximum rate is much higher.

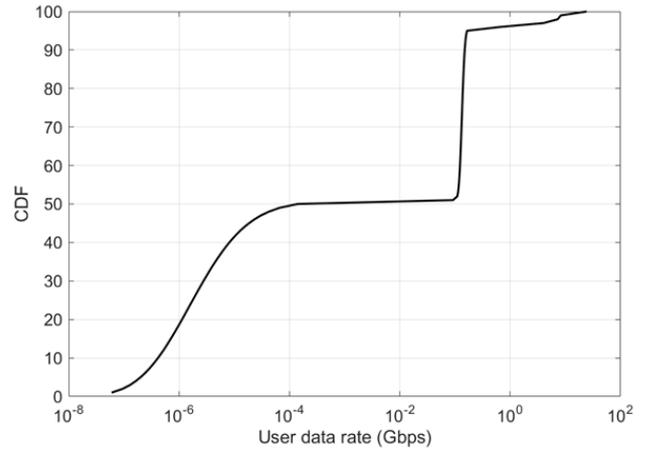

Fig. 4. Overall pdf of required data rate

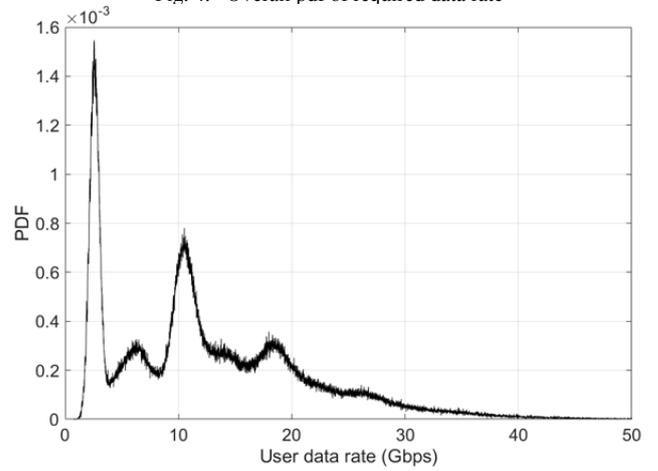

Fig. 5. Pdf of required data rate in a mm-wave cell

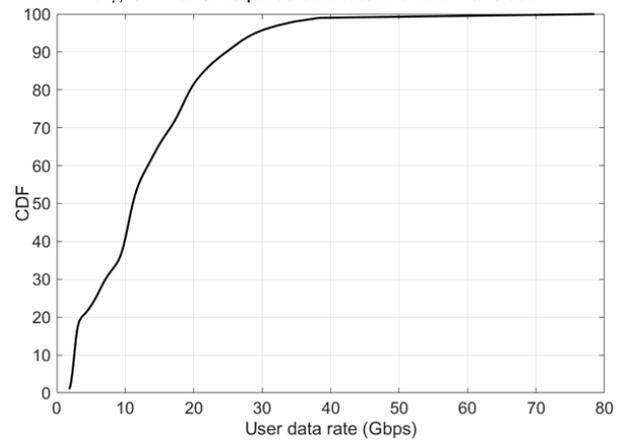

Fig. 6. CDF of required data rate in a mm-wave cell

*B. Overall Distribution Function*

The distribution of a single user with multiple traffic types is depicted in Fig. 4. It is shown that the data rate requirement can vary in a very wide range for such a user. The stepwise curve is caused by various rate requirements for different traffic type. The mean data rate is 318.4Mbps but the maximum data rate can be over 20Gbps. This distribution is in line with the baseline use case description in the sense that most of the users only requires 100M+ data rate but the peak rate requirement could be up to 20Gbps.

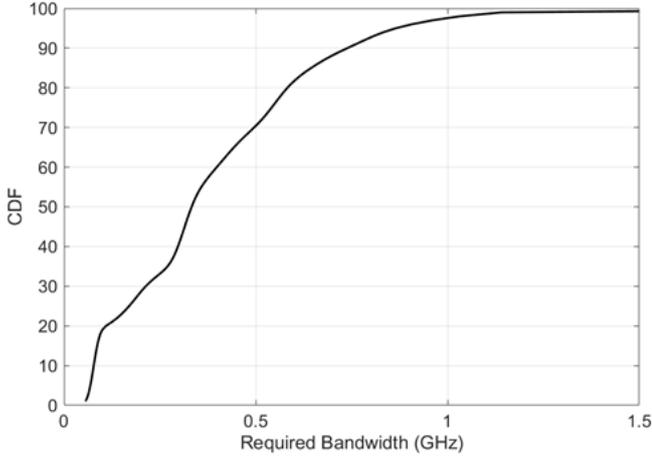

Fig. 7. CDF of required bandwidth in a mm-wave cell

If we assume there are multiple active users per each mm-wave small cell, the pdf and CDF of the aggregate required user rate are shown in Fig. 5 and Fig. 6, respectively. As seen from Fig. 5, there is a high peak at around 2.5 Gbps due to the aggregation of low data rate users for web browsing and content sharing. In addition, there are multiple low peaks above 5Gbps. The mean aggregate user rate is 12.75Gbps, i.e., about 40 times of an individual user. At least 29Gbps and 38Gbps data rates are needed in order to satisfy 95% and 99% users, respectively. The CDF of the required bandwidth is shown in Fig. 7. Based on the area spectral efficiency assumption denoted in (23) it can be seen that, with the above system parameters, the BW requirement is 1.15GHz to satisfy 99% users and 860MHz for 95% users. It is worth mentioning that for simplicity we assume orthogonal spectrum usage for all the users so that the spectral efficiency will not be degraded due to interference. The required BW could be even larger when non-orthogonal spectrum usage between users is considered.

## V. CONCLUSIONS

In this paper, we propose a new use case for 5G deployments, to satisfy the early adopters in 5G mobile data. The KPI requirements for the new use case are established and the aggregated data rates derived in combining 4 different potential traffic types. The statistical analysis in terms of plausible probability distributions for the traffic types lead to a cumulative distribution of the data rates, from which the bandwidth requirements are derived. Bandwidth results are highly dependent on the assumed average spectral efficiency, which in turn depends not only on the environment and channel model, but also on the foreseeable transmission techniques that may be in use for 5G. It is expected that this early work in a potential 5G resource requirement would encourage other work with more refined assumptions and would subsequently provide operators and other parties with a clearer view on the 5G bandwidth requirements.

## APPENDX A

If we let

$$K = \frac{1}{\sqrt{2\pi}T_{wb}\sigma\left(\Phi\left(\frac{\ln a_{up}-\mu}{\sigma}\right)-\Phi\left(\frac{\ln a_{low}-\mu}{\sigma}\right)\right)} \quad (A1)$$

The expression of $p_{Y,T}$ can be simplified as

$$p_{Y,T}(y,t) = K\exp\left(-\frac{(y-\mu)^2}{2\sigma^2}-\frac{1}{T_{wb}}t\right) \quad (A2)$$

Considering (8), we then have

$$p_{Y,T}(r_{wb}t,t) = K\exp\left(-\frac{(r_{wb}t-\mu)^2}{2\sigma^2}-\frac{1}{T_{wb}}t\right)$$

$$= K\exp\left(-\frac{\left(t-\left(\frac{\mu}{r_{wb}}-\frac{\sigma^2}{T_{wb}r_{wb}^2}\right)\right)^2+\left(\left(\frac{\mu}{r_{wb}}\right)^2-\left(\frac{T_{wb}r_{wb}\mu-\sigma^2}{T_{wb}r_{wb}^2}\right)^2\right)}{2(\sigma/r_{wb})^2}\right) \quad (A3)$$

$$= \overline{K}\frac{1}{\sqrt{2\pi}\overline{\sigma}}\exp\left(-\frac{(t-\overline{\mu})^2}{2\overline{\sigma}^2}\right), \quad \frac{\ln a_{low}}{r_{wb}} \le t \le \frac{\ln a_{up}}{r_{wb}}$$

where

$$\overline{K} = \sqrt{2\pi}\overline{\sigma}K\exp\left(-\frac{\left(\frac{\mu}{r_{wb}}\right)^2-\left(\frac{T_{wb}r_{wb}\mu-\sigma^2}{T_{wb}r_{wb}^2}\right)^2}{2(\sigma/r_{wb})^2}\right) \quad (A4)$$

$$\overline{\mu} = \left(\frac{\mu}{r_{wb}}-\frac{\sigma^2}{T_{wb}r_{wb}^2}\right), \overline{\sigma} = \sigma/r_{wb}$$

(A3) can be regarded as a truncated normal distribution. Therefore, (5) can be calculated as

$$f(r_{wb}) = \overline{K}\int_{\frac{\ln a_{low}}{r_{wb}}}^{\frac{\ln a_{up}}{r_{wb}}} t\frac{1}{\sqrt{2\pi}\overline{\sigma}}\exp\left(-\frac{(t-\overline{\mu})^2}{2\overline{\sigma}^2}\right)dt \quad (A5)$$

which is the mean of the truncated normal distribution (A3) scaled by $\overline{K}$, given as

$$f(r_{wb}) = \overline{K} \int_{\frac{\ln a_{low}}{r_{wb}}}^{\frac{\ln a_{up}}{r_{wb}}} t \frac{1}{\sqrt{2\pi}\overline{\sigma}} \exp\left(-\frac{(t-\overline{\mu})^2}{2\overline{\sigma}^2}\right) dt$$

$$= \overline{K}\left(\overline{\mu} + \frac{\overline{\sigma}\left(\varphi\left(\frac{(\ln a_{up})/r_{wb} - \overline{\mu}}{\overline{\sigma}}\right) - \varphi\left(\frac{(\ln a_{low})/r_{wb} - \overline{\mu}}{\overline{\sigma}}\right)\right)}{\Phi\left(\frac{(\ln a_{up})/r_{wb} - \overline{\mu}}{\overline{\sigma}}\right) - \Phi\left(\frac{(\ln a_{low})/r_{wb} - \overline{\mu}}{\overline{\sigma}}\right)}\right) \quad (A6)$$

## APPENDIX B

If we let

$$K = \frac{\alpha_x a_{x,low}^{\alpha_x}}{1 - \left(\frac{a_{x,low}}{a_{x,up}}\right)^{\alpha_x}} \frac{\alpha_t a_{t,low}^{\alpha_t}}{1 - \left(\frac{a_{t,low}}{a_{t,up}}\right)^{\alpha_t}} \quad (B1)$$

The expression of $p_{X,T}$ can be simplified as

$$p_{X,T}(y,t) = Kx^{-\alpha_x-1} t^{-\alpha_t-1} \quad (B2)$$

Considering (17), (B2) can be rewritten as

$$p_{X,T}(r_{uhd}t, t) = K(r_{uhd}t)^{-\alpha_x-1} t^{-\alpha_t-1}$$
$$= K r_{uhd}^{-\alpha_x-1} t^{-\alpha_x-1-\alpha_t-1} \quad (B3)$$

$$a_{\max} = \max\left\{\frac{a_{x,low}}{r_{wb}}, a_{t,low}\right\} \le t \le a_{\min} = \min\left\{\frac{a_{x,up}}{r_{wb}}, a_{t,up}\right\}$$

By letting

$$\overline{K} = \frac{K r_{uhd}^{-\alpha_x-1}\left(1 - \left(\frac{a_{\max}}{a_{\min}}\right)^{\alpha_x+\alpha_t+1}\right)}{(\alpha_x + \alpha_t + 1) a_{\max}^{\alpha_x+\alpha_t+1}} \quad (B4)$$

we have

$$p_{X,T}(r_{uhd}t, t) = \overline{K} \frac{\overline{\alpha} a_{\max}^{\overline{\alpha}} t^{-\overline{\alpha}-1}}{1 - \left(\frac{a_{\max}}{a_{\min}}\right)^{\overline{\alpha}}}, a_{\max} \le t \le a_{\min} \quad (B5)$$

which can be regarded as a truncated Pareto distribution. Therefore, (17) can be calculated as

$$f(r_{uhd}) = \overline{K} \int_{a_{\max}}^{a_{\min}} t \frac{\overline{\alpha} a_{\max}^{\overline{\alpha}} t^{-\overline{\alpha}-1}}{1 - \left(\frac{a_{\max}}{a_{\min}}\right)^{\overline{\alpha}}} dt \quad (B6)$$

Following the same calculation of web browsing traffic, (17) is given as

$$f(r_{uhd}) = \frac{\overline{K} a_{\min}^{\overline{\alpha}}}{1 - \left(\frac{a_{\max}}{a_{\min}}\right)^{\overline{\alpha}}} \left(\frac{\overline{\alpha}}{\overline{\alpha}-1}\right)\left(\frac{1}{a_{\min}^{\overline{\alpha}-1}} - \frac{1}{a_{\max}^{\overline{\alpha}-1}}\right) \quad (B7)$$

ACKNOWLEDGEMENTS

The European Commission funding under H2020- ICT-14-2014 (Advanced 5G Network Infrastructure for the Future Internet, 5G PPP), and project partners: Samsung, Ericsson, Aalto University, Alcatel-Lucent, CEA LETI, Fraunhofer HHI, Huawei, Intel, IMDEA Networks, Nokia, Orange, Telefonica, Bristol University, Qamcom, Chalmers University of Technology, Keysight Technologies, Rohde & Schwarz, TU Dresden are acknowledged.

REFERENCES

[1] The EU-5GPPP (H2020) funded mmMAGIC project: https://5g-mmmagic.eu/
[2] Nekovee M., et.al., "Millimetre-Wave Based Mobile Radio Access Network for Fifth Generation Integrated Communications (mmMAGIC)", Proc. of EuCNC, Paris, France, 29 June- 2 July, 2015.
[3] mmMAGIC D1.1, "Use case characterization, KPIs and preferred suitable frequency ranges for future 5G systems between 6 GHz and 100 GHz", available at https://5g-mmmagic.eu/results/#deliverables
[4] Samsung 5G vision white paper, Feb. 2015, available at: http://www.samsung.com/global/business-images/insights/2015/Samsung-5G-Vision-0.pdf
[5] BT Ultra HD review: http://www.techradar.com/reviews/audio-visual/digital-tv-receivers/bt-ultra-hd-youview-box-1301334/review.
[6] WiFi Alliance white paper, "WiGig® and the future of seamless connectivity", Sept. 2013, available at: http://www.wi-fi.org/download.php?file=/sites/default/files/private/WiGig_White_Paper_20130909.pdf
[7] 1xEV-DV Evaluation Methodology (V13), 2009
[8] Curtiss, J. H. (December 1941). "On the Distribution of the Quotient of Two Chance Variables". The Annals of Mathematical Statistics 12 (4): 409–421.
[9] 3GPP TR 36.814 v9.0.0, "Further advancements for E-UTRA physical layer aspects," 2010.
[10] R. M. Feldman and C. V. Flores, *Applied Probability and Stochastical Processes*, second edition, Springer, 2014.
[11] ITU-R report, "Recommendation ITU-R BT.2020," Aug. 2012.
[12] Online tutorial, "Bit depth tutorial".
[13] Yu Liu (2009-04-15). "The Preliminary Requirements for NGVC". H265.net. Retrieved 2012-11-25.
[14] T. Nakamura et.al., "Trends in small cell enhancements in LTE advanced", IEEE Comms Mag., Vol. 51 , No. 2, Feb. 2013.
[15] A. B. Carlson and P. B. Crilly, Communication Systems, McGraw-Hill Higher Education, 5 edition, 1 Mar. 2009.
[16] ITU-Report, "Future spectrum requirements estimate for terrestrial IMT", Dec. 2013.